\documentclass[a4paper,alpha-refs]{eSpectra}

\journal{aej}

\usepackage{graphicx}
\usepackage{siunitx}
\usepackage[spanish, english]{babel} 
\usepackage{amssymb}
\usepackage{comment}
\usepackage{amsmath}
\usepackage{xcolor}
\usepackage{subcaption}

\usepackage{listings}
\usepackage{xcolor}

\lstdefinestyle{py}{
  language=Python,
  basicstyle=\ttfamily\small,
  keywordstyle=\color{blue},
  commentstyle=\color{gray},
  stringstyle=\color{teal},
  showstringspaces=false,
  breaklines=true,
  frame=single,
  columns=fullflexible
}

\usepackage{academicons}
\usepackage{aas-macros}
\usepackage{doi}
\usepackage{fontawesome5}

\definecolor{orcidlogocol}{rgb}{0.65, 0.807, 0.223}
\newcommand{\orcid}[1]{$\,$\href{https://orcid.org/#1}{\textcolor{orcidlogocol}{\faOrcid}}}
 
\title{Identifying and Measuring Satellite Streaks in DECam Images}

\author[1,2,3\authfn{1}]{Alexandra Serrano Mendoza}
\author[4,2,3\authfn{2}]{Meredith L. Rawls}
\author[5,6,2,3\authfn{3}]{Andrés A. Plazas Malagón}

    \affil[1]{{Universidad Industrial de Santander, Bucaramanga, Colombia}}
    \affil[2]{{\textcolor{black}{IAU Centre for the Protection of the Dark and Quiet Sky, IAU-UAI Headquarters, Paris, France}}}
    \affil[3]{{RECA Internship Program}}
    \affil[4]{{University of Washington, USA}}
    \affil[5]{SLAC National Accelerator Laboratory, Menlo Park, CA, USA}
    \affil[6]{Kavli Institute for Particle Astrophysics and Cosmology, Stanford University, Stanford, CA, USA}

\authnote{\authfn{1} Undergraduate Student - Physics}
\authnote{\authfn{2} Research Scientist}
\authnote{\authfn{3} Rubin Operations Scientist (plazas@stanford.edu)}

\papercat{Scientific Article}

\runningauthor{Serrano Mendoza et al.}

\jvolume{4}
\jnumber{1}
\jyear{2026}
\begin{document}
\begin{frontmatter}
\maketitle

\selectlanguage{english}
\begin{abstract}
\justifying

The rapid growth of satellite constellations, particularly Starlink, is increasingly affecting ground-based astronomy. In this project, we developed a workflow to detect, identify, and measure the brightness of \textcolor{black}{trails from artificial satellites and other orbiting objects} in archival images from the Dark Energy Camera (DECam), available through the NOIRLab Data Archive. We filtered images with visible streaks, retrieved detector-level images, applied the Hough Transform (via satmetrics) to detect and align trails, and performed surface brightness photometry. We also used SatChecker to \textcolor{black}{obtain likely identifications} for each trail. \textcolor{black}{Our sample of nine measured streaks includes Starlink satellites, a navigation satellite, a decommissioned science satellite, and a rocket body.} Our results show that \textcolor{black}{satellites and other orbiting objects} are consistently detectable in DECam images, but their brightness varies significantly, reflecting design and operational differences across \textcolor{black}{object types and} models. While the methodology proved effective, detecting faint streaks was challenging, and short-lived glints remain an even harder problem for future work. This proof-of-concept establishes a foundation for larger statistical studies of satellite impacts on astronomical surveys. The code is available at \url{https://github.com/iausathub/reca-streaks}.
\end{abstract}


\selectlanguage{spanish} 
\begin{abstract}
\justifying
El rápido crecimiento de las constelaciones de satélites, en particular Starlink, está afectando cada vez más a la astronomía desde tierra. En este proyecto desarrollamos un flujo de trabajo para detectar, identificar y medir el brillo de posibles trazas de satélites Starlink en imágenes archivadas de la Cámara de Energía Oscura (DECam), disponibles a través del Archivo de Datos de NOIRLab. Filtramos imágenes con estelas visibles, recuperamos las imágenes a nivel de detector, aplicamos la Transformada de Hough (mediante satmetrics) para detectar y alinear las trazas, y realizamos fotometría de brillo superficial. También usamos SatChecker para obtener identificaciones probables de cada traza. Nuestros resultados muestran que los satélites Starlink son detectables de forma consistente en imágenes de DECam, pero su brillo varía significativamente, reflejando diferencias de diseño y operación entre modelos. Aunque la metodología resultó eficaz, la detección de estelas débiles fue desafiante, y los destellos breves representan un problema aún mayor para trabajos futuros. Este estudio de prueba de concepto establece las bases para análisis estadísticos más amplios sobre el impacto de los satélites en los estudios astronómicos. El código de este proyecto está disponible en \url{https://github.com/iausathub/reca-streaks}.
\end{abstract}

\begin{skeywords}
Fotometría, satélites
\end{skeywords}
\end{frontmatter}

\selectlanguage{english}
\section{Introduction}

In recent years, the night sky has changed noticeably. Since 2019, SpaceX alone has launched nearly 8000 Starlink satellites into orbit, and other companies have plans that could push the total number beyond 500,000 in the coming decade. Most of these satellites are placed in low-Earth orbit \textcolor{black}{(LEO)}, where they move faster and reflect sunlight, creating bright streaks in astronomical images. In astronomy, those lines can interfere with valuable data and critical observations. \textcolor{black}{Throughout this paper, we use the term ``satellite'' broadly to refer to any artificial object in Earth orbit; public catalogs of tracked objects include active and derelict satellites, rocket bodies, and space debris, all of which can produce trails in astronomical images.} The satellite constellation population is growing rapidly and it is therefore increasingly important to understand the satellite brightness distribution, its impact on science, and potentially necessary mitigations \citep{2023FallePaper,2022HuLSST,2021SATCON2Observations,2020TysonLSST}.

Although various studies have noted the presence of satellite trails and explored their impact on a specific telescope or scientific investigation \citep{2023LangHESS,2022MrozZTF}, one aspect remains relatively unexplored: cataloging individual observed trails with both the satellite ID and its brightness in a systematic way. The goal of this project is to demonstrate this is now possible and that it can help quantify the impacts satellites have on astronomy. In this work, we directly measure the \textcolor{black}{surface brightness} of 9 streaks in archival images from the Dark Energy Camera (DECam), mounted on the 4-meter Blanco telescope in Chile. \textcolor{black}{While our primary focus is on Starlink satellites given their dominance in the current LEO population, the workflow is general and our sample also includes non-Starlink objects such as a navigation satellite (NAVSTAR-70), a decommissioned science satellite (SORCE), and a rocket body (DELTA-2 R/B).}

This work contributes directly to the international efforts of SatHub at the IAU Centre for the Protection of the Dark and Quiet Sky (IAU CPS), which aims to understand and mitigate the impact of satellite constellations on astronomy. Specifically, once we have a set of identified satellites with measured brightnesses, it is possible to upload this information to SatHub's Satellite Constellation Observation Repository (SCORE). Data from multiple sources can then be used to understand how the number and brightness of these satellites have changed over the years.

In Section \ref{sec:methods}, we describe how we selected images with streaks, identified the \textcolor{black}{satellites}, and measured properties of each streak. In Section \ref{sec:results}, we present \textcolor{black}{our measurements (Table~\ref{tab:streak_photometry})} of each measured streak along with its likely \textcolor{black}{identification} and information about the corresponding DECam exposure and detector. In Section \ref{sec:discuss}, we discuss how our technique can be applied to more images and continue to improve our quantification of the impact of satellites on astronomy.


\begin{figure*}[ht]
    \centering
    \includegraphics[scale=0.35]{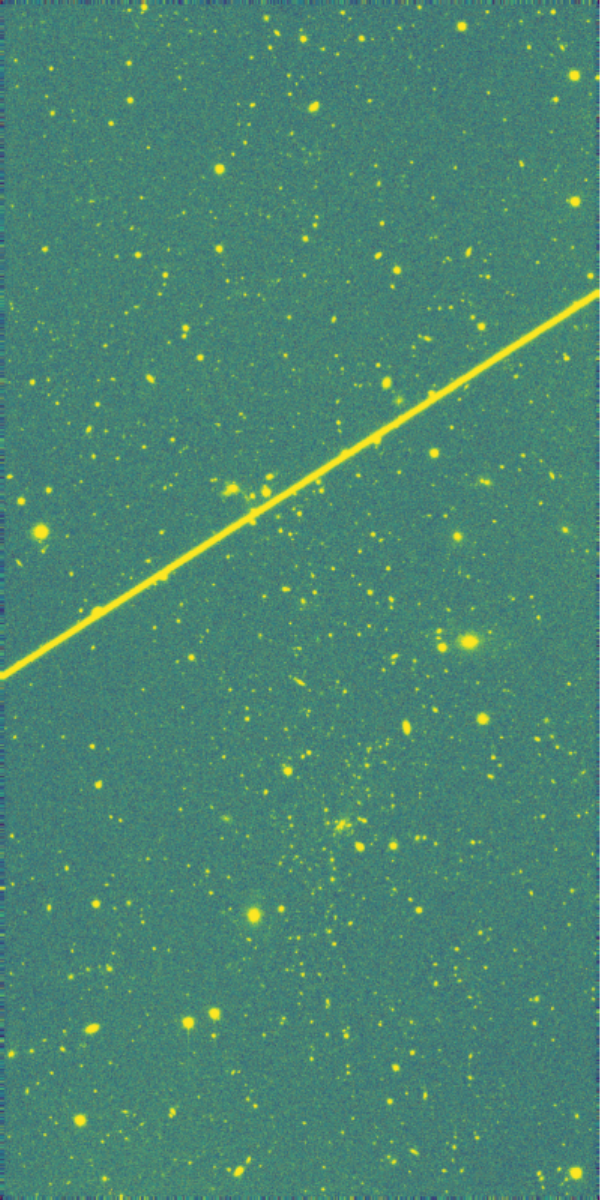}
    \includegraphics[scale=0.15]{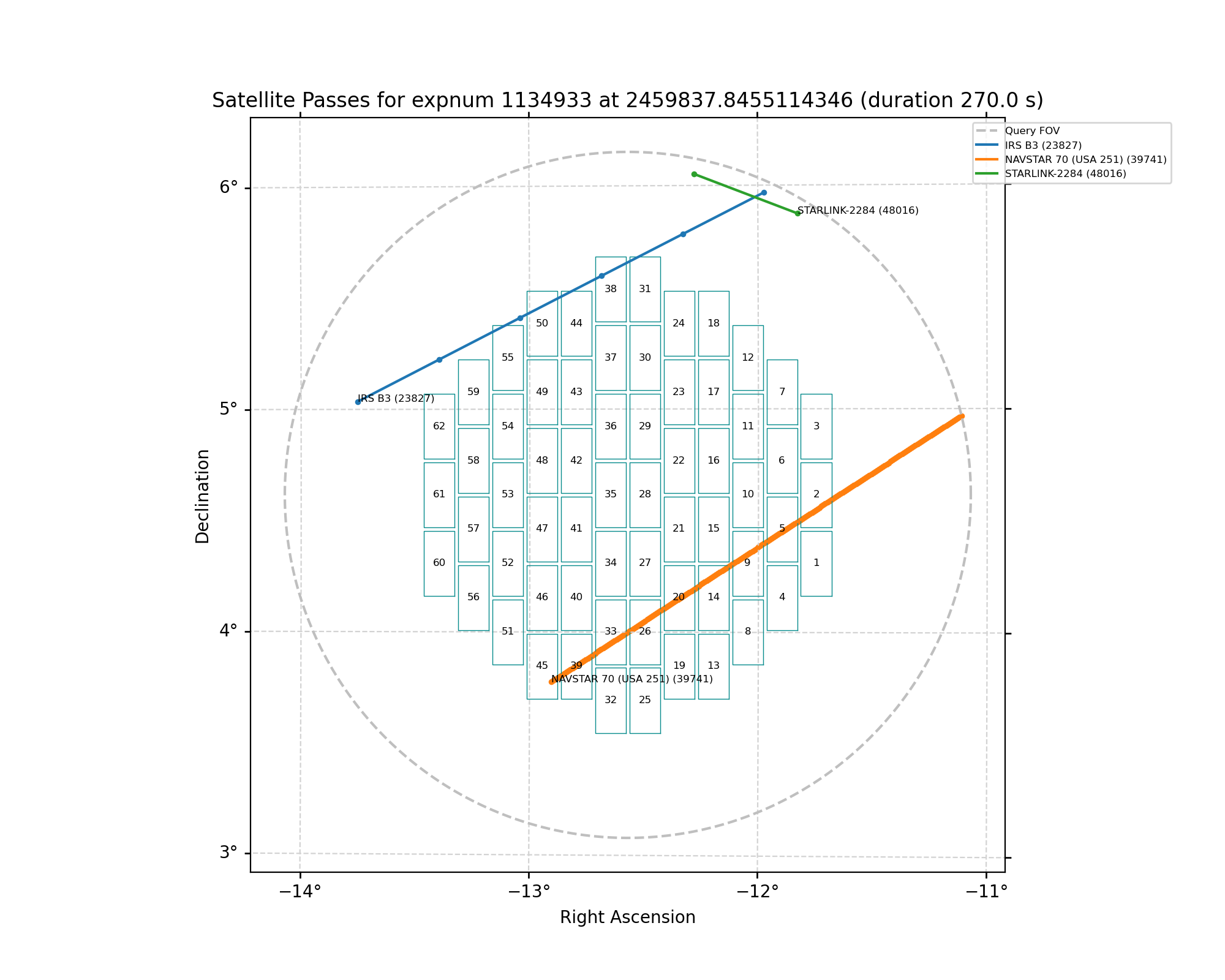}
    \caption{Example of a satellite trail in a DECam exposure \textcolor{black}{(left, CCD~5)} compared with SatChecker predictions for satellite passes (right) for NAVSTAR-70 in DECam exposure 1134933. \textcolor{black}{The left panel shows the individual CCD image containing the trail; the right panel shows the predicted satellite paths across the full DECam focal plane, with the trail of interest crossing CCD~5.} The predicted satellite path overlaps with the observed streak in the image, confirming the identification.}
    \label{fig:satcheck_comparison}
\end{figure*}

\section{Methods} \label{sec:methods}

This work focuses \textcolor{black}{primarily} on Starlink satellites, given their rapid growth and their leading impact on astronomical observations. Among the many existing constellations, Starlink is currently the most numerous and thus the most relevant for assessing the impact of satellite trails in optical astronomy, but there are also numerous inactive or leftover objects, such as rocket bodies, in Earth orbit that can leave comparably bright trails to those of active constellation satellites. \textcolor{black}{Our sample therefore includes both Starlink and non-Starlink objects to demonstrate the generality of the workflow.}

To study these effects, we used images from the Dark Energy Camera (DECam, \citet{flaugher2015}), a wide-field instrument mounted on the 4-meter Blanco Telescope at Cerro Tololo Inter-American Observatory (Chile) and publicly available through the NOIRLab Astro Data Archive.\footnote{\url{https://astroarchive.noirlab.edu/}} \textcolor{black}{DECam consists of 62 science CCDs arranged in a roughly hexagonal mosaic, covering a total field of view of approximately 3~square degrees. Each individual CCD has a field of view of roughly $9' \times 18'$ ($2048 \times 4096$~pixels at a plate scale of 0.263~arcsec/pixel). DECam is equipped with a set of broadband filters ($u$, $g$, $r$, $i$, $z$, $Y$) spanning the optical to near-infrared wavelength range ($\sim$300--1060~nm). For this proof-of-concept study, the images we analyze come from individual CCD extensions rather than full-mosaic exposures.} Since 2012, DECam has collected more than 7 million images, making it particularly sensitive to streaks caused by low Earth orbit satellites. This archive provides a unique opportunity to examine how the satellite population has evolved over time and how it affects ground-based observations.

The work presented here is based on datasets provided by A. Drlica-Wagner (private communication). These included a list of DECam exposures from the last several years and a catalog of corresponding candidate linear features that may be satellite trails. From this material, \textcolor{black}{we visually inspected images flagged as containing candidate linear features, checked whether those features could plausibly be Starlink satellite trails based on their morphology and brightness, and selected a small subset of exposures} with the aim of developing a workflow that combines trail detection, satellite identification, and brightness measurements for Starlink satellites or streaks with similar properties.

The workflow developed for this study was as follows:

\subsection*{Accessing detector images}
We initially selected a handful of DECam visits of interest based on visual inspection of a handful of images in A. Drlica-Wagner's datasets, described above. Images containing trails of interest were retrieved directly from the NOIRLab Astro Data Archive. Instead of downloading entire exposures, we accessed only the CCD extensions containing trails, optimizing both storage and analysis.

\subsection*{Streak detection and alignment}
To characterize the satellite trails, we used the Hough Transform, implemented in the {\tt{Satmetrics}} package.\footnote{\url{https://github.com/uwescience/satmetrics/tree/main}} This package detects linear features with configurable background thresholds, measures their angle in the image, and rotates them to a horizontal orientation \textcolor{black}{(see Figure~\ref{fig:hough})}, which simplifies and standardizes the subsequent photometric analysis. \textcolor{black}{The rotation is performed using an affine transformation via the OpenCV library with bilinear interpolation. We note that this resampling method is not strictly flux-conservative; however, for the purpose of surface brightness estimation at the level of precision relevant here (not millimag-level), the introduced error is small. Future work requiring higher photometric precision should extract fluxes directly from the unrotated images.} \textcolor{black}{The Hough Transform requires several tunable parameters, including the detection threshold, brightness cuts, and blur kernel size. We adopt default values (e.g., a threshold of 0.075 and a thresholding cut of 0.5) for most images, but found that some exposures---particularly those with faint streaks---required per-image parameter tuning to successfully detect the trail. This manual tuning is a limitation for scaling the workflow to large datasets, which we discuss further in Section~\ref{sec:discuss}. See Appendix \ref{app:pipeline_repro} for a list of the parameters used}. 

\subsection*{Streak identification with SatChecker}

\textcolor{black}{To determine the likely origin of each trail, we used the SatChecker FOV tool\footnote{\url{https://satchecker.readthedocs.io/en/latest/}} \citep{2024DadighatCPS}. Given the sky coordinates and observation time of a DECam exposure, SatChecker propagates archival two-line element (TLE) orbital data for cataloged objects and identifies which satellites are predicted to cross the instrument field of view during the exposure. The predicted trail positions and orientations are then overlaid on the focal plane, as illustrated in Figure~\ref{fig:satcheck_comparison}. We visually compared the predicted paths to the observed streaks and selected the best-matching satellite identification. In most cases the match was unambiguous, particularly for exposures containing a single bright trail. For exposure 1103448, which contained four Starlink streaks from satellites in the same orbital shell, SatChecker predicted four closely spaced passes and the matching relied on the relative positions of the trails across different CCDs.}

Because SatChecker relies on archival TLE data, its predictions are not guaranteed to be accurate. We do not measure \textcolor{black}{quantitative} positional offsets between visually identified streak locations and SatChecker forecasted streak locations in this work, but in the future, such measurements would help validate SatChecker and improve its utility for either avoiding or intentionally observing certain satellites.

\subsection*{Photometric measurements}

\textcolor{black}{Finally, we performed surface brightness photometry on the identified trails using both aperture photometry and forward-model fitting. We measured the surface brightness in units of counts per square pixel and, when a photometric zeropoint was available from the DECam instcal (instrument-calibrated) data product, we also report values in AB~mag/arcsec$^2$. The instcal zeropoints are stored in the \texttt{MAGZERO} FITS header keyword and are derived from the DECam community pipeline calibration process; they correspond to the AB magnitude system. For exposures lacking a valid zeropoint, we report results only in counts per square pixel. We note that because the exposures in our sample were taken in different filters and on different CCDs (with potentially different quantum efficiencies), the count-based measurements are not directly comparable across entries in Table~\ref{tab:streak_photometry}.}

\subsubsection{Aperture Photometry Method}

The first approach to estimate the surface brightness of satellite streaks is a
non-parametric, aperture-based method. The image is \textcolor{black}{first} pre-processed to mask
\textcolor{black}{contaminating bright point sources (e.g., stars) whose pixel values} exceed a user-defined sigma threshold,
\(\texttt{median} + N_\sigma \times \texttt{std}\), where \(N_\sigma\) is typically 5.
\textcolor{black}{These masked pixels are excluded from subsequent calculations so that the streak flux
is not biased by overlapping stars. The remaining unmasked pixels---which include both
the streak and the sky background---are used to construct} a 2D masked array from which a 1D cross-trail
intensity profile is computed as the mean flux along the trail direction.

The profile is modeled with a Gaussian function:
\begin{equation}
I(y) = A \exp\left[-\frac{(y - y_0)^2}{2\sigma^2}\right] + B,
\end{equation}
where \(A\) is the amplitude, \(y_0\) is the centroid of the streak, \(\sigma\) is
the Gaussian width in pixels, and \(B\) is the background offset.
The full width at half maximum (FWHM) of the streak cross-section is then \(\mathrm{FWHM} = 2.355 \, \sigma\).

After identifying the on-streak region within \(\pm 3\sigma\) of the fitted
centroid, two symmetric off-streak regions of equal width are used to estimate
the background level \(B\). The total streak flux, after background subtraction, is:
\begin{equation}
F_\mathrm{streak} = \sum_{\mathrm{on}} (I - B),
\end{equation}
where the summation is taken over the unmasked pixels in the aperture.

To convert the total flux to a mean surface brightness in physical units,
the aperture area in arcsec$^2$ is computed as:
\begin{equation}
A_\mathrm{aper} = L_\mathrm{trail} \times 3\mathrm{\sigma} \times s^2,
\end{equation}
where \(L_\mathrm{trail}\) is the trail length in pixels, $\mathrm{\sigma}$ is the
fitted Gaussian width in pixels, and \(s\) is the DECam plate scale ($0.263$ arcsec/pixel).

The surface brightness in counts per arcsec$^2$ is:
\begin{equation}
\mu_\mathrm{counts} = \frac{F_\mathrm{streak}}{A_\mathrm{aper}}.
\end{equation}
If a photometric zeropoint \(ZP\) is available from the DECam instcal data product, the corresponding surface
brightness in mag/arcsec$^2$ is:
\begin{equation}
\mu_\mathrm{mag} = ZP - 2.5 \log_{10} \mu_\mathrm{counts}.
\end{equation}

The flux uncertainty is computed assuming independent noise contributions
from the sky background, read noise, and Poisson shot noise:
\begin{equation}
\sigma_F^2 = 
\begin{cases}
N_\mathrm{pix} (\sigma_B^2 + \sigma_R^2), & \text{if background-dominated}, \\
F_\mathrm{e^-} + N_\mathrm{pix} (\sigma_B^2 + \sigma_R^2), & \text{if source-dominated},
\end{cases}
\end{equation}
where \(F_\mathrm{e^-}\) is the signal in electrons, \(\sigma_B\) is the standard
deviation of the background, \(\sigma_R\) is the read noise (both in electrons),
and \(N_\mathrm{pix}\) is the number of pixels in the on-streak aperture.
The surface brightness uncertainty is then:
\begin{equation}
\sigma_\mu = \frac{\sigma_F}{A_\mathrm{aper}}.
\end{equation}

This method is straightforward and robust for bright or isolated streaks,
as it makes minimal assumptions about the streak’s underlying profile.
However, because the Gaussian fit is only applied to a 1D cross-trail slice,
flux losses can occur in the wings of the point-spread function (PSF), especially for faint or
extended streaks. The background estimation may also be biased if spatial
gradients or residual stars are present near the trail.

\subsubsection{PSF-Convolved Trail Fitting Method}

A more accurate approach is to model the streak as a point source moving
linearly during the exposure, convolved with a Gaussian point-spread function.
\textcolor{black}{This model is fitted to the rotated (horizontally aligned) trail image, but
retains the angle parameter \(\theta\) to account for any residual orientation
offset after rotation.}
Following \citet{Veres2012}, the two-dimensional trail model is:

\begin{equation}
\begin{split}
f_T(x, y) = b + \frac{\Phi}{L} \cdot 
\frac{1}{2\sigma\sqrt{2\pi}}\,
\exp\!\left[
-\frac{((x - x_0)\sin\theta + (y - y_0)\cos\theta)^2}
{2\sigma^2}
\right]
\\[4pt]
\times
\left\{
\mathrm{erf}\!\left(\frac{Q + L/2}{\sigma\sqrt{2}}\right)
-
\mathrm{erf}\!\left(\frac{Q - L/2}{\sigma\sqrt{2}}\right)
\right\},
\end{split}
\label{eq:trail-profile}
\end{equation}
where \(b\) is the background level, \(\Phi\) is the total flux,
\(L\) is the trail length in pixels, \(\sigma\) is the PSF width,
\(\theta\) is the trail orientation angle, and \((x_0, y_0)\) is the
trail center. The variable \(Q\) corresponds to the coordinate along
the trail:
\begin{equation}
Q = (x - x_0)\cos\theta + (y - y_0)\sin\theta.
\end{equation}

\begin{figure*}[ht]
  \centering

  \begin{subfigure}[t]{0.49\textwidth}
    \centering
    \includegraphics[width=0.5\linewidth]{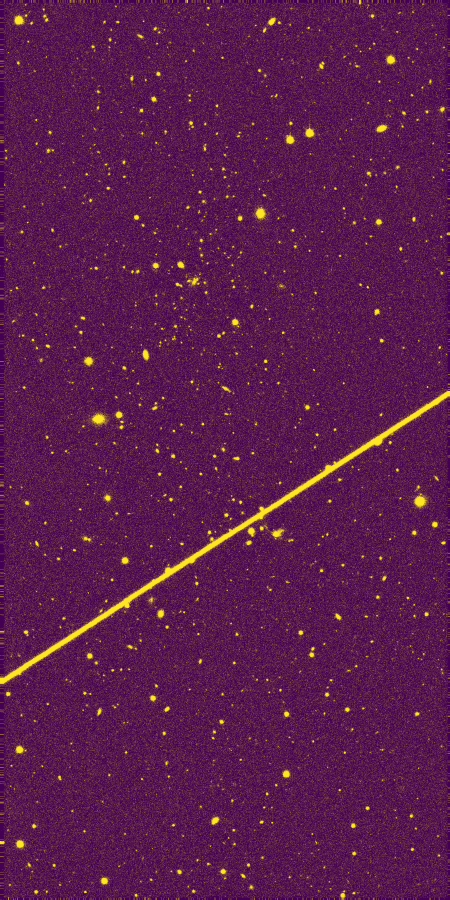}
  \end{subfigure}
  \hfill
  \begin{subfigure}[t]{0.49\textwidth}
    \centering
    \includegraphics[width=0.5\linewidth]{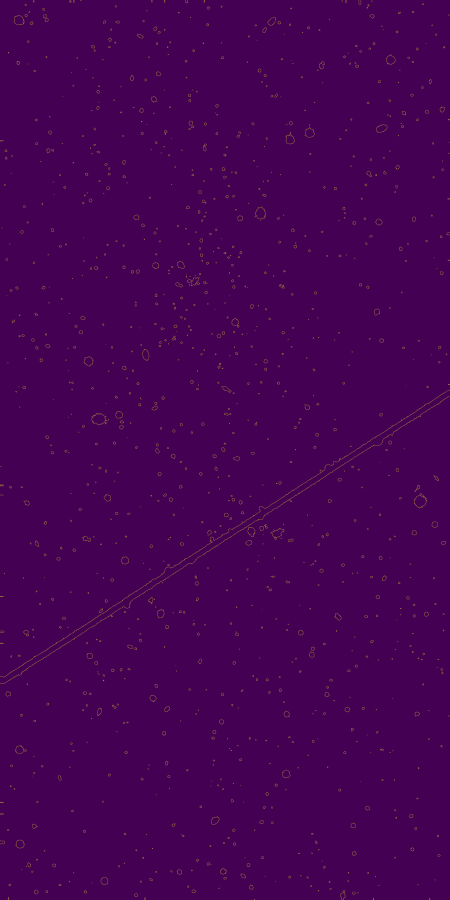}
  \end{subfigure}

  \vspace{0.5ex}

  \begin{subfigure}[t]{\textwidth}
    \centering
    \includegraphics[width=\linewidth]{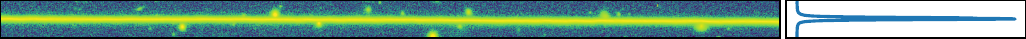}
  \end{subfigure}

  \caption{\textcolor{black}{Detection and alignment of a satellite trail using the Hough Transform (via \texttt{satmetrics}). Top left: binary thresholded image produced by the line detection algorithm, highlighting candidate linear features above the background. Top right: Canny edge-detected image, which is used by the Hough Transform to identify and measure the orientation of the trail. Bottom: extracted 2D cutout of the trail after rotation to a horizontal orientation, with the median cross-trail intensity profile overlaid. Pixel values are in counts (ADU).}}
  \label{fig:hough}
\end{figure*}

The model parameters are fitted using non-linear least-squares
minimization using \texttt{scipy.optimize.curve\_fit}.
The total flux \(\Phi\) and background \(b\) are obtained directly from the fit.

The mean surface brightness is then derived analogously to the aperture method:
\begin{equation}
\mu_\mathrm{counts} = \frac{\Phi}{A_\mathrm{fit}},
\quad A_\mathrm{fit} = L \times 3\mathrm{\sigma} \times s^2,
\end{equation}

The flux uncertainty \(\sigma_\Phi\) is estimated from the covariance matrix
of the fit, scaled by the root-mean-square (RMS) of the residuals:
\begin{equation}
\sigma_\Phi = \sqrt{C_{11}} \, \sigma_\mathrm{res},
\end{equation}
where \(C_{11}\) is the variance of \(\Phi\) from the covariance matrix.
\textcolor{black}{Because the fit residuals incorporate all noise sources present in the data
(sky background, read noise, and Poisson shot noise), these contributions are
implicitly included in $\sigma_\Phi$ through $\sigma_\mathrm{res}$, unlike
the aperture method where they are modeled explicitly (Equation~6).}
The corresponding uncertainty in surface brightness is
\(\sigma_\mu = \sigma_\Phi / A_\mathrm{fit}\).

When a photometric zeropoint \(ZP\) is available, the surface brightness
in magnitudes per arcsecond squared is computed as:
\begin{equation}
\mu_\mathrm{mag} = ZP - 2.5 \log_{10} \mu_\mathrm{counts},
\end{equation}
and the associated magnitude uncertainty is given by the logarithmic propagation
of flux error:
\begin{equation}
\sigma_{\mu, \mathrm{mag}} = \frac{1.0857}{\mathrm{SNR}},
\end{equation}
where the numerical constant \(1.0857 = 2.5 / \ln 10\).
The signal-to-noise ratio (SNR) is computed following Equation (8) of
\citet{Veres2012}:
\begin{equation}
\mathrm{SNR} = \frac{S}{\sqrt{S + B}},
\end{equation}
where \(S = \Phi\) is the total fitted streak signal and
\(B = b \cdot N_\mathrm{pix}\) is the background contribution
within the model aperture.

\textcolor{black}{This SNR formulation follows \citet{Veres2012} and accounts for Poisson noise
from both the signal and sky background, which are the dominant noise sources
for the bright streaks in our sample; read noise is a subdominant contribution
that is not included in this simplified expression.}
The approximation
\(\sigma_{\mu, \mathrm{mag}} = 1.0857 / \mathrm{SNR}\) is valid for
high signal-to-noise ratios (\(\mathrm{SNR} \gtrsim 10\)), where the
fractional flux error is small and the magnitude error distribution
remains symmetric.

\begin{figure*}[th]
    \centering
    \includegraphics[width=0.75\linewidth]{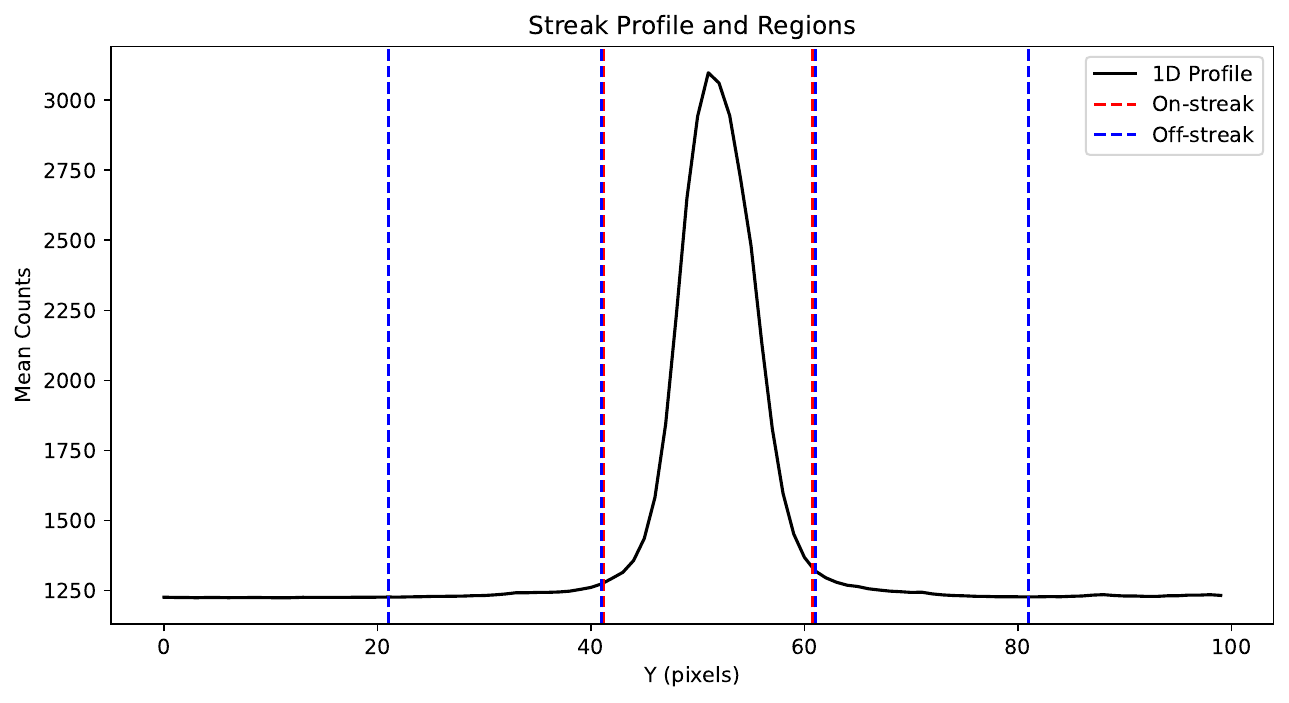}\\[0.6em]
    \includegraphics[width=0.75\linewidth]{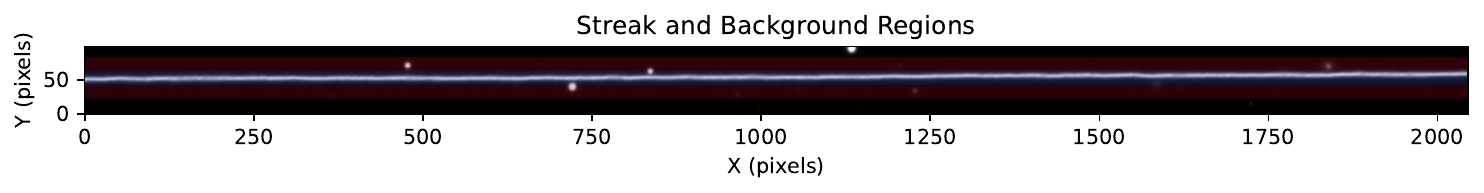}
    \caption{\textcolor{black}{Photometric analysis of a satellite trail (NAVSTAR-70, exposure 1134933, CCD~5; see Table~\ref{tab:streak_photometry}). Top: 1D brightness profile (in counts) across the streak (mean across the long axis), showing a clear signal above the background. The on-streak and off-streak (background) regions are indicated with dashed vertical lines. Bottom: corresponding 2D image with the regions used for signal and background measurement.}}
    \label{fig:photometry}
\end{figure*}

\section{Results} \label{sec:results}
The workflow described above allowed us to successfully detect and measure the brightness of several satellite trails in DECam images. The combination of SatChecker predictions \textcolor{black}{(Figure~\ref{fig:satcheck_comparison})}, Hough Transform line detection \textcolor{black}{(Figure~\ref{fig:hough})}, and photometric analysis \textcolor{black}{(Figure~\ref{fig:photometry})} proved effective for isolating and characterizing individual streaks. \textcolor{black}{Our measurements are summarized in Table~\ref{tab:streak_photometry}.}

First, the Hough Transform in satmetrics worked well to find the orientation of the trails. By rotating them to be horizontal, we could compare them more easily and prepare them for photometry. This step was useful because trails appear in many different directions across the DECam focal plane.

\textcolor{black}{The photometric measurements show significant variability in brightness across our sample. The brightest streak (NAVSTAR-70, a GPS satellite) has a surface brightness of $\sim$18.7~mag/arcsec$^2$, while the faintest Starlink streak (STARLINK-2559) measures $\sim$24.8~mag/arcsec$^2$. This range spans more than 6 magnitudes and reflects differences in object type, size, orbital altitude, attitude, and illumination geometry.}

\textcolor{black}{The two photometric methods generally yield consistent results, with typical agreement better than $\sim$0.2~mag.}

\textcolor{black}{We note that the formal uncertainties reported in Table~\ref{tab:streak_photometry} are statistical only and are likely underestimated, particularly for the highest-SNR objects. For example, the $\pm$0.0002~mag/arcsec$^2$ uncertainty for NAVSTAR-70 from the fitting method reflects only the formal fit covariance and does not account for systematic effects such as flat-fielding errors, background gradients, zeropoint calibration uncertainty, or the motion of the satellite during the exposure. Realistic total uncertainties are likely at the level of a few millimagnitudes or larger.}

Even though we only worked with a small number of images, these results confirm that the method is valid. The workflow can \textcolor{black}{detect and identify linear features from various types of orbiting objects}, measure their brightness, and point to trends that deserve more detailed study with larger datasets.

\begin{table*}[ht]
\centering
\caption{Surface brightness of satellite streaks measured from DECam exposures. 
Columns list the satellite name, NORAD catalog identifier (from the N2YO satellite database at {\url{https://www.n2yo.com/database/}}), exposure number, CCD detector, surface brightness (SB) measured via aperture photometry and trail fitting (both in mag~arcsec$^{-2}$ when a photometric zeropoint was available, or counts~arcsec$^{-2}$ otherwise), signal-to-noise ratio (SNR) from the Section~3 definition, and photometric zeropoint (ZP) in magnitudes. The DECam plate scale of 0.263~arcsec~pix$^{-1}$ is adopted throughout.}
\label{tab:streak_photometry}
\renewcommand{\arraystretch}{1.05}
\setlength{\tabcolsep}{3pt}
\small
\begin{tabular}{lccccccccc}
\hline
\textbf{Name} & \textbf{NORAD ID} & \textbf{Expnum} & \textbf{Det} &
\textbf{SB$_{\text{Aper}}$} & \textbf{SB$_{\text{Fit}}$} &
\textbf{SNR} & \textbf{ZP} &
\textbf{SB$_{\text{Aper}}$} & \textbf{SB$_{\text{Fit}}$} \\
 &  &  &  & (mag/arcsec$^2$) & (mag/arcsec$^2$) &
  & (mag) & (counts/arcsec$^2$) & (counts/arcsec$^2$) \\
\hline

NAVSTAR-70 & 39741 & 1134933 & 5  &
$19.198 \pm 0.001$ & $18.79 \pm 0.0002$ &
4509.5 & 30.051 &
21943.17 $\pm$ 10.89 & 31885.38 $\pm$ 0.00 \\

STARLINK-2600 & 48387 & 1138498 & 23 &
$23.551 \pm 0.020$ & $23.47 \pm 0.0036$ &
301.9 & 31.006 &
959.51 $\pm$ 17.41 & 1034.91 $\pm$ 6872.67 \\

STARLINK-2559 & 48298 & 1033925 & 17 &
$24.962 \pm 0.016$ & $24.84 \pm 0.0275$ &
39.5 & 29.450 &
62.38 $\pm$ 0.93 & 70.10 $\pm$ 12.79 \\

STARLINK-3758 & 52297 & 1103448 & 4 &
-- & -- &
1386.5 & -- &
2562.46 $\pm$ 2.25 & 2777.78 $\pm$ 0.00 \\

STARLINK-3772 & 52290 & 1103448 & 20 &
-- & -- &
1440.1 & -- &
2854.92 $\pm$ 1.16 & 3200.09 $\pm$ 0.00 \\

STARLINK-3771 & 52293 & 1103448 & 45 &
-- & -- &
1272.3 & -- &
2215.26 $\pm$ 1.67 & 2654.65 $\pm$ 47.96 \\

STARLINK-3765 & 52295 & 1103448 & 34 &
-- & -- &
1419.9 & -- &
2906.02 $\pm$ 6.03 & 2986.22 $\pm$ 0.00 \\

DELTA-2 R/B & 20763 & 1072590 & 5 &
— & — &
1906.7 & — &
12520.96 $\pm$ 7.39 & 15479.47 $\pm$ 0.00 \\

SORCE & 27651 & 1125268 & 21 &
— & — &
36.4 & — &
8.71 $\pm$ 1.35 & 20.75 $\pm$ 10.73 \\

\hline
\end{tabular}
\end{table*}

\section{Discussion} \label{sec:discuss}


This project showed that it is possible to detect and measure satellite trails in DECam images, but it also highlighted several challenges. One of the main difficulties was identifying faint streaks. Sometimes they were clear to the eye, but harder for the automatic detection tools to capture, which means parameter adjustments were necessary. This is important because newer Starlink models and many other satellites are designed to be dimmer, so working with faint signals will only become more common.

\textcolor{black}{Several aspects of the current workflow require manual intervention that would need to be automated or standardized to scale to large datasets. First, the Hough Transform detection parameters had to be tuned on a per-image basis for faint streaks, as noted in Section~\ref{sec:methods} (see appendix \ref{app:pipeline_repro}). Second, the satellite identification via SatChecker currently relies on visual matching of predicted and observed trail positions rather than an automated cross-match with quantitative offset tolerances. Addressing these limitations through automated parameter optimization and programmatic SatChecker matching would be necessary steps toward a large-scale survey of satellite trail brightnesses.}

Another challenge we did not fully address here is the presence of glints, brief sudden flashes caused by satellite surfaces reflecting sunlight \textcolor{black}{\citep{2023KarpovZTF}}. Unlike trails, which extend across the image, glints are short and localized, making them harder to predict and to separate from stars or image artifacts. Dealing with glints will likely require new strategies and represents the next step after trails.

Despite these challenges, the workflow combining SatChecker, Hough Transform, and photometry worked well for a proof-of-concept. Going forward, the same approach could be scaled to larger datasets and applied to other instruments, building a broader statistical picture of how satellites affect astronomy.

\textcolor{black}{An important extension of this work would be to compute apparent magnitudes in addition to surface brightnesses. Since we have the TLE data for each identified object, it is in principle possible to compute the time the satellite spends crossing the detector (the streak exposure time) and thereby convert the surface brightness to an apparent magnitude. This would enable direct comparison with brightness measurements from other sources such as SCORE \citep{2024RawlsSatHub}, and is left for future work.}

The code is available at \url{https://github.com/iausathub/reca-streaks}.

\section{Acknowledgments}
We thank Alex Drlica-Wagner for essential past work and assisting with key parts of the project. We thank Siegfried Eggl for valuable conversations that greatly improved the project. \textcolor{black}{We also thank the two anonymous reviewers for constructive comments that improved this paper.}
\textcolor{black}{The authors acknowledge the support of the International Astronomical Union (IAU) Centre for the Protection of the Dark and Quiet Sky (CPS). The Centre coordinates collaborative and multidisciplinary international efforts from institutions and individuals working across multiple geographic areas, seeks to raise awareness, and mitigate the negative impact of satellite constellations on ground-based optical, infrared and radio astronomy observations as well as on humanity's enjoyment of the night sky. Any opinions, findings, and conclusions or recommendations expressed in this material are those of the author(s) and do not necessarily reflect the views of the IAU, NSF NOIRLab, SKAO, ESO, or any host or member institution of the IAU CPS.}
The development of SatChecker and SCORE has been supported by the National Science Foundation (NSF) under grant number AST 2332736. SatChecker and SCORE are hosted at the NSF NOIRLab. This research uses services or data provided by the Astro Data Archive at NSF NOIRLab. NOIRLab is operated by the Association of Universities for Research in Astronomy (AURA), Inc. under a cooperative agreement with the NSF.
The work of AAPM was supported by the U.S. Department of Energy under contract number DE-AC02-76SF00515. AAPM thanks the Department of Physics of Harvard University and the Laboratory of Particle Astrophysics and Cosmology, the Cosmology Group at Boston University, and the Department of Physics at Washington University in St. Louis for their hospitality during the preparation of this paper.
This research was conducted as part of the RECA (Network of Colombian Astronomy Students) Internship \textcolor{black}{\citep{reca2026,reca2023}} Program 2025.

\section{Contributions}
ASM carried out the analysis and wrote most of the manuscript. MLR conceived of the project, assisted with the analysis, and wrote some of the manuscript. AAPM served as link with RECA, assisted with the project and analysis as well as writing and editing the manuscript.

\appendix
\section{Validation of Streak Photometry via Simulations}
\label{app:simulation-validation}

To validate the accuracy and error estimation of our two methods for measuring the surface brightness (SB) of satellite streaks—(1) rectangular aperture photometry and (2) 2D trail profile fitting—we performed controlled simulations using the \texttt{GalSim} library \citep{rowe2015}.

\subsection{Simulation Setup}

We simulate a PSF-convolved trail with known flux and geometry, then measure its surface brightness using both methods and compare the recovered values to the true inputs \textcolor{black}{(Table~\ref{tab:sim-results})}.

The trail is modeled as a box profile with:
\begin{itemize}
    \item Total flux: $F_\mathrm{input} = 2.5 \times 10^6$ ADU
    \item Trail length: $L = 2048$ pixels $= 540.06$ arcsec
    \item Trail width: $w = 1$ pixel (before convolution)
    \item Trail angle: $\theta = 0^\circ$
\end{itemize}

The trail is convolved with a circular Gaussian point-spread function (PSF) of full width at half maximum:
\[
\mathrm{FWHM}_\mathrm{PSF} = 0.9~\mathrm{arcsec}
\]

The pixel scale is set to:
\[
s = 0.263~\mathrm{arcsec/pixel}
\]

We include a uniform background with:
\[
B = 3800~\mathrm{counts}/\mathrm{arcsec}^2 \Rightarrow 264.24~\mathrm{counts/pixel}
\]

Read noise is added at a level of $6~e^-$ RMS, and the detector gain is set to $g = 3.95~e^-$/ADU. The image is rendered in a $100 \times 2048$ pixel array.

The true surface brightness is calculated as:

\begin{multline}
\mathrm{SB}_\mathrm{input} = 
\frac{F_\mathrm{input}}{L_\mathrm{arcsec} \times \mathrm{FWHM}_\mathrm{PSF}} 
= \frac{2.5 \times 10^6}{540.06 \times 0.9} \\
= 5143.48 \pm 2.17~\mathrm{counts/arcsec}^2
\end{multline}

where the error is propagated assuming Poisson noise from the signal and background, and including read noise.

\subsection{Recovered Values}

\begin{table}[ht]
\centering
\caption{Recovered Surface Brightness from Simulated Trail}
\label{tab:sim-results}
\begin{tabular}{lcccc}
\hline
\textbf{Method} & \textbf{Flux} & \textbf{SB} & \textbf{Error} & \textbf{SNR} \\
& \textbf{(ADU)} & \textbf{(ADU/arcsec$^2$)} &  &  \\
\hline
Input (true)     & $2.50 \times 10^6$ & 5143.48 & 2.17 & --- \\
Fitting (Eq.~\ref{eq:trail-profile})  & $2.50 \times 10^6$ & 4961.56 & 19.08 & 1157.7 \\
Aperture         & $2.49 \times 10^6$ & 4935.05 & 2.88 & 1713.4 \\
\hline
\end{tabular}
\end{table}

Both methods recover the true surface brightness within $\sim 4\%$ of the input value. The aperture method provides tighter errors due to the stricter definition of the integration region, but may underestimate the total flux when wings of the trail profile are excluded. The fitting method, while more model-dependent, accounts for the convolution with the PSF and can provide robust estimates even in cases of faint trails.

Both methods are suitable for estimating the surface brightness of satellite trails in DECam images. The aperture method is fast and relatively assumption-free, while the PSF-convolved model fitting approach provides more accurate flux estimates when the trail shape and PSF are well-described by the model.

\section{Streak detection and photometry pipeline: wrapper function, usage example, and tuned parameters}
\label{app:pipeline_repro}

This work combines line detection and image rotation utilities from the \texttt{satmetrics} repository
and photometry tools from the \texttt{reca-streaks} repository.
Line detection and image rotation are provided by \texttt{satmetrics} (see repository:
\texttt{https://github.com/iausathub/satmetrics}), while the photometry routines developed here are in
\texttt{reca-streaks} (\texttt{https://github.com/iausathub/reca-streaks}).

\subsection{Wrapper function for Hough-based line detection}
\label{app:wrapper}

\begin{lstlisting}[style=py,caption={Wrapper function used to configure and run \texttt{satmetrics} Hough-based line detection on DECam images.},label={lst:detect_lines_hough}]
def detect_lines_hough(
    image,
    threshold=0.075,
    flux_prop_thresholds=[0.1, 0.2, 0.3, 1],
    blur_kernel_sizes=[3, 5, 9, 11],
    brightness_cuts=(2, 2),
    thresholding_cut=0.5,
    **kwargs
):
    """
    Detect linear streaks in a 2D astronomical image via the satmetrics Hough pipeline.
    """
    lineDetector = ld.LineDetection(image)
    lineDetector.threshold = threshold
    lineDetector.flux_prop_thresholds = flux_prop_thresholds
    lineDetector.blur_kernel_sizes = blur_kernel_sizes
    lineDetector.brightness_cuts = brightness_cuts
    lineDetector.thresholding_cut = thresholding_cut

    for key, value in kwargs.items():
        setattr(lineDetector, key, value)

    detections = lineDetector.hough_transformation()
    return detections
\end{lstlisting}

\subsection{Example usage in the full pipeline}
\label{app:usage}

\begin{lstlisting}[style=py,caption={Example pipeline usage combining \texttt{satmetrics} (line detection and rotation) with \texttt{reca-streaks} photometry routines.},label={lst:pipeline_usage}]
import sys
from astropy.io import fits

# Local clones (used in this work)
sys.path.append('/PATH/TO/satmetrics')
sys.path.append('/PATH/TO/reca-streaks')

import line_detection_updated as ld
import image_rotation as ir
import streak_photometry

# Example inputs
hdulist = fits.open(file_name)
image_data = hdulist[1].data

brightness_cuts = (3, 5)
thresholding_cut = 0.08

# 1) Detect candidate lines (wrapper around satmetrics LineDetection)
detected_lines = streak_photometry.detect_lines_hough(
    image_data,
    brightness_cuts=brightness_cuts,
    thresholding_cut=thresholding_cut
)

# 2) Cluster detected lines (satmetrics)
clustered_lines = ld.cluster(
    detected_lines["Cartesian Coordinates"],
    detected_lines["Lines"]
)

# 3) Rotate image so the best line is horizontal (satmetrics)
rotated_images, best_fit_params = ir.complete_rotate_image(
    clustered_lines=clustered_lines,
    angles=detected_lines["Angles"],
    image=image_data,
    cart_coord=detected_lines["Cartesian Coordinates"]
)

stripe = rotated_images[0]

# 4) Photometry on the rotated stripe (reca-streaks)
streak_photometry.streak_photometry_aperture(
    stripe, hdu_list=hdulist, make_plots=make_plots
)
streak_photometry.streak_photometry_psf_fitting(
    stripe, hdu_list=hdulist, make_plots=make_plots
)
\end{lstlisting}

\subsection{Hand-tuned parameters for reproducibility}
\label{app:handtuned}

Table~\ref{tab:handtuned_satmetrics_all} lists the parameter values used for the Hough-based detection
(wrapper in Listing~\ref{lst:detect_lines_hough}). Unless otherwise stated, the wrapper defaults were used:
$\texttt{threshold}=0.075$, $\texttt{flux\_prop\_thresholds}=[0.1,0.2,0.3,1]$,
$\texttt{blur\_kernel\_sizes}=[3,5,9,11]$, $\texttt{brightness\_cuts}=(2,2)$,
$\texttt{thresholding\_cut}=0.5$.

\begin{table*}[ht]
\centering
\caption{Line-detection wrapper parameters used for each satellite streak measurement (matching the satellites listed in Table~\ref{tab:streak_photometry}). Entries labeled ``defaults'' indicate the wrapper defaults; otherwise we list the manual overrides.}
\label{tab:handtuned_satmetrics_all}
\setlength{\tabcolsep}{5pt}
\renewcommand{\arraystretch}{1.1}
\small
\begin{tabular}{lccccp{7.8cm}}
\hline
\textbf{Name} & \textbf{NORAD ID} & \textbf{Expnum} & \textbf{Det} & \textbf{ZP available?} & \textbf{Wrapper parameters used} \\
\hline
NAVSTAR-70      & 39741 & 1134933 & 5  & Yes  & defaults \\
STARLINK-2600   & 48387 & 1138498 & 23 & Yes  & defaults \\
STARLINK-2559   & 48298 & 1033925 & 17 & Yes  & $\texttt{brightness\_cuts}=(3,5)$; $\texttt{thresholding\_cut}=0.08$ \\
STARLINK-3758   & 52297 & 1103448 & 4  & No   & $\texttt{brightness\_cuts}=(3,5)$; $\texttt{thresholding\_cut}=0.08$ \\
STARLINK-3772   & 52290 & 1103448 & 20 & No   & $\texttt{brightness\_cuts}=(3,5)$; $\texttt{thresholding\_cut}=0.08$ \\
STARLINK-3771   & 52293 & 1103448 & 45 & No   & $\texttt{brightness\_cuts}=(3,5)$; $\texttt{thresholding\_cut}=0.08$ \\
STARLINK-3765   & 52295 & 1103448 & 34 & No   & $\texttt{brightness\_cuts}=(3,5)$; $\texttt{thresholding\_cut}=0.08$ \\
DELTA-2 R/B     & 20763 & 1072590 & 5  & No   & defaults \\
SORCE           & 27651 & 1125268 & 21 & No   & defaults \\
\hline
\end{tabular}
\end{table*}

\clearpage


\bibliography{biblio}

\end{document}